\documentclass[a4paper,fleqn]{cas-sc}
\usepackage{amsmath}
\usepackage{autobreak}
\usepackage{amsfonts,amssymb}
\usepackage{xcolor}
\usepackage{ulem}
\usepackage[numbers,sort&compress]{natbib} 

\DeclareMathOperator*{\argmin}{arg\,min}

\usepackage{mathtools}

\def\bfs{{\bf s}}

\newcommand{\bbeta}{\mbox{\boldmath $\beta$}}

\usepackage{siunitx}

\begin{document}
\let\WriteBookmarks\relax
\def\floatpagepagefraction{1}
\def\textpagefraction{.001}
\shortauthors{Yin et~al.}
\shorttitle{Risk Based Arsenic Rational Sampling Design}

\title [mode = title]{Risk Based Arsenic Rational Sampling Design for Public and Environmental Health Management}


\author[1,2]{Lihao Yin}
\cormark[1]
\address[1]{Department of Statistics, Texas A\&M University, College Station, Texas, 77843, USA}
\address[2]{Institute of Statistics and Big Data, Renmin University, Beijing, China}

\author[1]{Huiyan Sang}
\cormark[1]
\author[3]{Douglas J. Schnoebelen}
\cormark[2]
\author[3 ]{Brian Wels}
\address[3]{University of Iowa, Iowa City, IA 52242}
\author[4 ]{Don Simmons}
\address[4]{State Hygienic Laboratory, University of Iowa, Coralville, Iowa, 52241, USA}
\author[4 ]{Alyssa Mattson}
\author[4]{Michael Schueller}
\author[4 ]{Michael Pentella}
\author[5]{Susie Y. Dai} 
\address[5]{Department of Plant Pathology and Microbiology, Texas A\&M University, College Station, TX, 77843, USA}
\cormark[3]

\cortext[cor1]{Equally contributing authors.}
\cortext[cor2]{Current contact: U.S. Geological Survey, 5563 De Zavala Road, San Antonio, TX 78023}
\cortext[cor3]{Corresponding author. Department of Plant Pathology and Microbiology, College Station, TX, 77843, USA. E-mail address: sydai@tamu.edu (Susie Dai PhD).}






\begin{abstract}
Groundwater contaminated with arsenic has been recognized as a global threat, which negatively impacts human health. Populations that rely on private wells for their drinking water are vulnerable to the potential arsenic-related health risks such as cancer and birth defects. Arsenic exposure through drinking water is among one of the primary arsenic exposure routes that can be effectively managed by active testing and water treatment. From the public and environmental health management perspective, it is critical to allocate the limited resources to establish an effective arsenic sampling and testing plan for health risk mitigation. We present a spatially adaptive sampling design approach based on an estimation of the spatially varying underlying contamination distribution. 
The method is different from traditional sampling design methods that often rely on a spatially constant or smoothly varying contamination distribution. 
In contrast, we propose a statistical regularization method to automatically detect spatial clusters of the underlying contamination risk from the currently available private well arsenic testing data in the USA, Iowa. This approach allows us to develop a sampling design method that is adaptive to the changes in the contamination risk across the identified clusters. 
 We provide the spatially adaptive sample size calculation and sampling location determination at different acceptance precision and confidence levels for each cluster. The spatially adaptive sampling approach may effectively mitigate the arsenic risk from the resource management perspectives. The model presents a framework that can be widely used for other environmental contaminant monitoring and sampling for public and environmental health.  
\end{abstract}


\maketitle


\section{Introduction}

Arsenic (As) is ranked as the 20th most abundant element in the Earth’s crust and has been studied internationally. Groundwater contaminated with arsenic has been recognized as a global threat, negatively impacting human health \citep{podgorski2020global,desimone2009quality}. The primary human exposure to arsenic is drinking water with additional contributors such as food and air \cite{almberg2017arsenic,vahter2009effects,sohel2009arsenic}. Arsenic is a potent human carcinogen, which can cause bladder, lung, and skin cancers \citep{argos2012arsenic}.  Furthermore, arsenic and its metabolites can cross the placental barrier and create risk for adverse maternal and fetal health, leading to adverse birth outcomes \citep{bloom2014maternal}. The Environmental Protection Agency (EPA) federal drinking water standard established 0.01 mg/L as the arsenic maximum contaminant levels (MCLs) in drinking water. In the USA, approximately 41.8 million (13\% of the total US population) people obtain drinking water from private wells, and the private wells are not regulated under the current EPA regulation \citep{national2016groundwater}. The recent national Water-Quality Assessment Program from the United States Geological Survey (USGS) reports that more than one out of five wells contain contaminants at concentrations exceeding the EPA MCLs or USGS health-based screening levels. Among the various pollutants that exceed the EPA maximum contaminant levels, arsenic contamination is a common finding. Because private wells are not regulated in the US, in the Midwest region, a significant percentage of the population depending on private wells for drinking water is at risk due to drinking water arsenic contamination \citep{schnoebelen2017elevated}. Arsenic testing in private well water represents a fundamental mean that helps mitigate the arsenic risk in the rural population for public and environmental health. In reality, many of the private wells are not tested, which presents a significant challenge for health risk mitigation. From the management perspective, a scientifically sound sampling plan to test a representative sample size is needed to characterize the environmental arsenic hazard with limited resources.


Sampling theory can be used to guide a large number of chemical and biological analyses for environmental control and consumer safety \cite{minkkinen2004practical}. As for arsenic testing,  a systematic sampling plan is critical for risk assessment to draw science and data-based conclusions and make the best usage of limited resources. The EPA has published guidance for data quality objectives with regard to sampling design \citep{us2002guidance}. One of the key preparations for a sampling design is to determine the sample size and sampling error for representative sample collection. 

Understanding sample statistical distributions is critical when selecting a sampling method, sampling strategy, and sample size. Application of probability distribution can help develop a science-based sampling plan and estimate the chemical and biological hazards in the environment. 
Previously, binomial probability theory has been well studied for sample size determination for estimating a binomial proportion \cite{gonccalves2012sample}.
Application examples include the sampling plan in product inspection and surveillance \citep{lee2016application}, epidemiology \citep{sepulveda2015sample}, and medical diagnostics \citep{joseph2005statistical}.  In many of these applications, a univariate binomial distribution is considered, that is, the underlying binomial proportion parameter is assumed to be a constant in the study.  However, due to the spatial heterogeneity nature of arsenic distribution in the earth's crust and groundwater, the traditional binomial sampling scheme based on a univariate binomial distribution may not be suitable to survey the target private well population. There is a great need to develop new sampling schemes capable of accounting for the spatially heterogeneity nature of the arsenic distribution. 

In terms of arsenic contamination, quite a few statistical and mathematical models have been used to estimate and predict arsenic concentrations in groundwater and private wells. Logistic models for binomial distributions are widely adopted to estimate the spatial distribution of As contamination probability at both global and regional levels \cite{amini2008statistical,ayotte2006modeling,winkel2008predicting,podgorski2017extensive,winkel2011arsenic,rodriguez2013groundwater,yang2012can}.  For instance, a logistic linear regression model has been used to predict the high arsenic domestic well population in the US \citep{ayotte2017estimating}. Furthermore, boosted regression tree models (weak-learner ensemble models) and traditional logistic linear models have been compared to estimate and predict arsenic contamination probabilities in drinking water wells in the Central Valley, California \citep {ayotte2016predicting}. Similar to those statistical models, predictive variables are used to predict geogenic arsenic in drinking water wells
in glacial aquifers, north-central USA \cite{erickson2018predicting}. Machine learning models have also been used to predict arsenic concentrations in groundwater in Asia \citep{tan2020machine}. Nevertheless, the aforementioned models primarily focus on the estimation and prediction of arsenic distributions rather than the sampling design. Moreover, most methods often rely on a rich set of predictors and training data set to guarantee model accuracy. 
To the best of our knowledge, there is very limited work that combines the model-based estimation of varying arsenic distributions with the binomial sampling  design method.

To close this gap in the current literature for spatial binomial distribution sampling design, the current study proposes a spatially adaptive sampling design approach, by estimating a spatially clustered underlying contamination distribution. We apply this method to determine the data locations to understand arsenic contamination in private wells in Iowa. The method is different from traditional spatial sampling design methods \citep{zhu2006spatial,diggle2010geostatistical} that often assume continuous process-based spatial models for relatively smooth spatial fields. In contrast, we model the underlying contamination risk as a spatially clustered function for a straightforward interpretation of the result. It also has the advantage of detecting discontinuous spatial heterogeneity in the arsenic distribution and then borrowing information within each identified spatially homogeneous cluster for an adaptive sampling design.
The method is built upon a graph fused lasso regularization method \citep{tibshirani2005sparsity}, which automatically detects clusters of spatial units and estimates the underlying spatially varying contamination distributions simultaneously. Thanks to the flexibility of graphs, our spatial clustering model enjoys several nice properties. First, it leads to very flexible cluster shapes naturally satisfying spatial contiguity constraints. Second, the method automatically learns the number of clusters from the data, relaxing the limitation in other clustering algorithms that require to specify the number of clusters a priori. 

Another unique advantage of estimating a spatially clustered contamination distribution over other contamination distribution estimation methods lies in its easy integration with the traditional binomial sampling theory.  Within each identified spatial cluster, the contamination distribution can be treated as having a common binomial proportion, for which we propose and compare two different sample size determination methods at different levels of acceptance precision and confidence. Given the sample size calculations, a remaining sample design task is to determine the sampling locations. In our study, both the candidate wells and the available tested wells are distributed highly unevenly in the study region. To ensure the sampling design has a balanced spatial coverage, we propose a practical algorithm based on spatial point processes to distinguish areas that have been sufficiently-sampled and insufficiently-sampled, and  determine new sampling locations accordingly. This new strategy, presumably more adaptive than traditional sampling without considering heterogeneity in sampling distributions, can potentially provide more precise tools to efficiently allocate sample collection efforts and resources.     

\section{Materials and Methods}
\subsection{Sample Collection and Analysis}\label{sec:method1} 
For the private well samples, the data used to build the model was collected as part of the Iowa Grants-to-Counties (GTC) program. The Iowa GTC program was established in 1987 after the Iowa legislature passed the Iowa Groundwater Protection Act to protect groundwater. Arsenic testing has been included as part of the GTC program based on Iowa Administrative Code \citep{Iowa}. A total of $14,570$ samples were collected and analyzed at the University of Iowa State Hygienic Laboratory from July 1st, 2015 to June 16, 2020. As part of the GTC program, the local health department collects the private well samples by conducting a home visit, and sending them to a laboratory for analysis. It should be noted that the selection of the laboratory is at the county's discretion. 

For all the samples analyzed at the State Hygienic Laboratory, the water sample is collected either at the tap faucet or outside the house. Samples are collected in a 4 oz. HDPE plastic bottle containing 1 mL of $1+1$ nitric acid as a preservative.  Cooling is not required for sampling.  Samples are screened for turbidity following Standard Methods 2130 B using a HACH model 2100N Turbidimeter.  Samples exceeding 1 nephelometric turbidity units (NTU) are digested prior to analysis. The arsenic analysis is performed based on the Iowa State Hygienic Laboratory standard operating procedure (SOP), similar to the EPA 200.2 method. Briefly, a 50-mL aliquot is transferred from a well-mixed sample to a polypropylene digestion tube (Environmental Express \#UC475-GN).  One mL of 1+1 nitric acid and 0.5 mL of $1+1$ hydrochloric acid (Fisher, Trace Metal Grade) are added to the tubes.  Digestion is accomplished using a hot block (Environmental Express \#SC154) at approximately $85^{\circ}$C.  The sample volume is reduced to 10 mL, and then the sample is covered with a watch glass (Environmental Express \#SC505), and refluxed for 30 minutes.  The tubes are cooled and diluted to 25 mL with reagent water.  The samples are further diluted to 50 mL using a mixture of 2\% nitric acid and 1\% hydrochloric acid.
The samples are then analyzed for arsenic using an Agilent 7500 CE inductively coupled plasma mass spectrometer following EPA method 200.8.  Approximately 5 mL of sample is transferred to a polypropylene autosampler tube for analysis.  The instrument is calibrated using a multi-point calibration curve (0, 1, 5, 50, 100, 500 ug/L).  Standards are matrix-matched to the sample.  Thus, digested samples are not analyzed in the same run with direct analysis samples.  Internal standards are introduced via a mixing tee at the instrument.  Yttrium is used as the internal standard for arsenic.  Results are not reported unless all quality controls pass their acceptance limits per the method.

The raw data amount to $14,570$ previously collected observations of Arsenic tests in total (Figure 1). Based on the risk categories, we characterize the wells that contain higher than 0.01 mg/L arsenic as high risk wells, and use a binary variable to denote whether a well is at high risk. We exclude the observations whose location information is absent. 
We also aggregate the repeated measurements at the same locations into one single observation, by setting the binary value to be $1$ if there is at least one concentration measurement exceeding MCL. A visual presentation of the private well arsenic testing is available through the Iowa Department of Public Health website \citep{url2}.  

\subsection{Estimation of spatially clustered contamination probabilities}\label{sec:method2}

Let $y(\bfs_i)$ denote the binary variable at a well location $\bfs_i$, for $i=1,\ldots, n$, coded as being $1$ if the arsenic concentration at $\bfs_i$ is exceeding the EPA MCL (i.e., 0.01 mg/L), and $0$ otherwise. Here, $n$ is the total number of available tested wells. 
We propose a spatially varying binary logistic model for $y(\bfs)$. Specifically, we assume
\begin{equation}\label{eq:mod1}
    P\big(y(\bfs_i)=1\big)\sim \text{Bernoulli}\big(p(\bfs_i)\big), \quad \text{for } i=1,\ldots, n, 
\end{equation}
where $p(\bfs_i)$ is the probability of the well located at $\bfs_i$ being contaminated.  
In the logistic regression model, we model the probability $p(\bfs_i)$ as
\begin{equation*}
    p(\bfs_i)=\frac{1}{1+\exp\{-\beta(\bfs_i)\}}
\end{equation*}
or equivalently,  $\log \frac{p(\bfs_i)}{1-p(\bfs_i)}=\beta(\bfs_i)$,
where $\beta(\bfs_i)$ is interpreted as the log-odds of the arsenic contamination event that $y(\bfs_i)=1$.  
Let $\bbeta=\big(\beta(\bfs_1),\ldots,\beta(\bfs_n) \big)$ be the stacked regression parameters for all the observed well locations. It follows that the corresponding logistic regression log likelihood function takes the form: 
\begin{equation}\label{eq:ell}
    \ell(\bbeta)=-\sum_{i=1}^{n}\log (1+e^{\beta(\bfs_i)})+\sum_{i=1}^{n} y(\bfs_i)\beta(\bfs_i)
\end{equation}

It is noted from \eqref{eq:mod1} that we relax the assumption of having a constant contamination probability $p$, or equivalently, contamination log-odds, $\beta$, over the whole study region, and instead assume it is varying over space. This assumption is reasonable for a large study region like Iowa due to the anticipated spatial heterogeneity in the arsenic concentration in groundwater and private wells. Specifically, we assume $p(\bfs)$ is a spatially clustered function, that is, there exists a number of geographical clusters such that $p(\bfs)$ stays relatively homogeneous within each cluster but varies across clusters. This will facilitate the easy visualization and interpretation of the varying contamination probability across different identified clusters.  We will show in Section \ref{sec:sample} that the spatially clustered contamination probability estimation also leads to an efficient  spatially adaptive sampling design strategy. 

We consider a flexible regularization model for pursing the clustered pattern of $\beta(\bfs)$ and $p(\bfs)$. Regularization methods have gained large popularity in modern high dimensional statistics and machine learning methods for various statistical learning tasks \citep{buhlmann2011statistics}. They have proved to be effective in imposing structural assumptions on model parameters such as sparsity, smoothness, and clustering to avoid over-fitting problems. The regularization method for the Arsenic contamination model is performed in the following steps:
\begin{enumerate} 
\item Construct a spatial graph, denoted as $G=(V,E)$ where $V=\{{v_1,v_2,...,v_n}\}$ is the vertex set with $n$ vertices and $E$ is the edge set. For a spatial problem, each vertex represents a spatial location. For example, in the arsenic case study, each vertex $v_i$ represents a well location $\bfs_i$, and the edge set $E$ reflects the prior assumption on the neighborhood structure of well locations based on spatial proximity. The edge set selection is an important component of the method, which we will discuss later in this section. 
\item Use the graph from step 1 to construct a homogeneity pursuit regularization, also called the fused lasso penalty function \citep{tibshirani2005sparsity,tibshirani2011solution} , for $\bbeta$ as follows:
\begin{equation}\label{eq:penalty}
\rho\sum_{(i,j)\in{E}} |\beta(\bfs_i)-\beta(\bfs_j))|.
\end{equation}
\item Combine the penalty function in \eqref{eq:penalty} with the logistic log-likelihood function in \eqref{eq:ell} to form a penalized objective function, which we minimize to obtain an estimator of $\bbeta$ as follows:  
\begin{equation}\label{eq4}
\hat{\bbeta}=\argmin_{\bbeta}Q(\bbeta)=\argmin_{\bbeta}\{-\frac{1}{n}\ell(\bbeta)+\rho\sum_{(i,j)\in E}|\beta(\bfs_i)-\beta(\bfs_j)|\}.
\end{equation}
\item After obtaining $\hat{\bbeta}$, calculate the estimate of the contamination probability from $\hat{p}(\bfs_i)=\dfrac{1}{1+\exp(-\hat{\beta}\big(\bfs_i)\big)}$.
\end{enumerate}
The fused lasso regularization in step 2 is used to impose the assumption that the arsenic contamination probabilities at two wells are more likely to take the same value if they are connected by an edge in $E$ of the specified spatial graph. The objective function $Q(\bbeta)$ in \eqref{eq4} takes a similar form as the standard negative log-likelihood function from Bernoulli distributions for binary arsenic data, but with an added fused lasso regularization term to encourage spatial clustering of $\bbeta$. 
As a result, when estimating the arsenic contamination probabilities from this penalized objective function $Q(\bbeta)$, we not only use the information from the binary arsenic testing data in the first likelihood term, but also take into account the spatial information from the spatial-graph based fused lasso penalty in the second term. $\rho$ is a regularization tuning parameter determining the strength of fused lasso penalty and ultimately influencing the estimated number of clusters of wells. 
The solution of $L_1$ norm penalty results in an exact fusion or separation between $\beta(\bfs_i)-\beta(\bfs_j)$, that is, the edges in the graph are classified into 
two sets, one consists of all the non-zero elements of $\beta(\bfs_i)-\beta(\bfs_j)$ corresponding to pairs of neighboring wells that have different contamination probabilities, and the other set consists of all the zero elements of $\beta(\bfs_i)-\beta(\bfs_j)$ corresponding to pairs of neighboring wells that share the same contamination probability. As such, this regularization automatically leads to spatially clustered contamination probabilities.

The choice of graph plays two important roles in the method; it not only reflects the prior information about the geological topology and spatial clustering constraint of the data, but also determines the computation complexity of the algorithm. Some natural graph choices for spatial data include the $k$ nearest neighbor graphs, graphs connecting neighbors within a certain radius, and spatial Delaunay triangulation graphs (see, e.g., \citet{li2019spatial}). Alternatively, graphs can be constructed based on some preliminary estimates of parameters. For instance, the differences between the initial estimates of parameters at any two vertices can be used as the distance metric between vertices to replace the spatial Euclidean distance when constructing graphs. In this paper, we take a hybrid approach to construct the graph; the $k$ nearest neighbor edge set connecting counties is determined based on the sample proportion within each county, and the $k$ nearest neighbor edge set within each county is determined based on the Euclidean distance. 

There are several advantages of using the fused lasso penalty function for cluster detection. First, this penalization allows to detect clusters and estimate model parameters simultaneously. Second, 
this method guarantees to achieve a spatially contiguous clustering configuration such that only adjacent locations are clustered together. Another appealing property of this method is that the resulting clusters have very flexible shapes. We explain this point using the notion of connected components in graph theory; spatially contiguous clusters can be defined as the connected components of a graph $G$, and accordingly, a spatially contiguous partition of $V$ can be defined as a collection of disjoint connect components such that the union of vertices is $V$.
It is easy to show that any spatially contiguous partition with arbitrary cluster shapes can be recovered by removing a set of edges from a spatial graph \citep{li2019spatial}. In addition, the number of clusters does not need to be fixed a priori. Instead, we can determine it by a data-driven information criterion approach described later in this section. Finally, besides its capability to capture piece-wise constant coefficients, previous theoretical studies proved that this penalty has a strong local adaptivity in that it is also capable of capturing piece-wise Lipschitz continuous functions \citep{madrid2020adaptive}, which implies that the method can also approximate a spatially smoothly varying contamination probability reasonably well.  
 
We now discuss how to solve the optimization in \eqref{eq4} to obtain the parameter estimation results. Note that $-\frac{1}{n}\ell(\bbeta)$ is convex and differentiable with respect to $\bbeta$, and $\sum_{(i,j)\in E}|\beta(\bfs_i) - \beta(\bfs_j)|$ is also convex.  
Therefore we propose an iterative algorithm combining the proximal gradient method \citep{beck2009fast} and the alternating direction method of multipliers (ADMM) \citep{boyd2011distributed} for this convex optimization problem. Specifically, given the current estimate $\bbeta^{(t)}$, we let $\mathbf{g}^{(t)} = \bbeta^{(t)} + (1/L)\frac{1}{n}\nabla{\ell}(\bbeta^{(t)})$, where $L$ is the Lipschitz constant of $-\frac{1}{n}\ell(\bbeta)$, and $\nabla{\ell}(\bbeta^{(t)})$ is the first derivative of $\ell(\bbeta)$ evaluated at $\bbeta^{(t)}$. 
For the logistic regression model in \eqref{eq:ell}, we can choose $L$ to be $1/n$. Following the proximal gradient algorithm, we then update the value of $\bbeta$ by solving:
\begin{equation}\label{Eq:max_f}
       \bbeta^{t+1} = \argmin_{\bbeta}   \frac{1}{2}\left\Vert \bbeta - \mathbf{g}^{(t)}\right\Vert_{2}^2 + \dfrac{\rho}{L}\sum_{(i,j)\in E}|\beta(\bfs_i)-\beta(\bfs_j)|.
\end{equation} 
We use the ADMM algorithm \citep{wahlberg2012admm} to solve the optimization in \eqref{Eq:max_f}. We will release the R code of our algorithm as a supplementary file upon acceptance of this manuscript for publication.  

Finally, the parameter estimation algorithm involves the selection of the tuning parameter $\rho$. In high dimensional statistics, data-dependent model selection criteria, such as generalized cross-validation  \citep{golub1979generalized}, Bayesian information criterion (BIC) \citep{schwarz1978estimating} and extended Bayesian information criterion \citep{chen2008extended} have been commonly used to determine the value of $\rho$. For the numerical studies in this paper, we use BIC with the form,  
$\texttt{BIC}=-2\ell(\hat{\beta})+k\log n$, where $k$ is the estimated number of clusters. The ``optimal" $\rho$ is selected by minimizing BIC from a candidate set.
 
\subsection{Spatially adaptive sampling design}\label{sec:sample}
We now turn the attention to the sampling design problem for the determination of the sample size and sample locations of wells.
Recall in Section~\ref{sec:method2} we have obtained a spatially clustered contamination probability $p(\bfs)$, that is, within each identified spatial cluster, each sample is assumed to have the same probability of being contaminated.  
This allows us to employ existing sampling design methods based on the univariate binomial distribution with a constant p within each cluster, while adapting to the value of $p$ across clusters.  
The method leads to a simple but efficient sampling strategy accounting for the spatial variation in $p(\bfs)$. 

Sample size determination and confidence interval construction methods for a constant-proportion binomial distribution have been well studied in the statistics literature.  
Popular methods include the Clopper-Pearson exact method, Wilson score method, Wald test, Bayesian Jeffreys method, and Agresti–Coull method, among others. 
For a review and comparison of different methods, see, for example, \cite{newcombe1998two} and \cite{gonccalves2012sample}. 
In this work, we consider two methods, the modified Jefferey and the Wilson score methods, following the recommendations by \cite{gonccalves2012sample}.

Consider a univariate binomial distribution where a random sample of size $n$ is drawn from a large population, $X$ is the number of 1's (e.g., the number of contaminated wells), and $p$ is the probability of a randomly selected well is contaminated. We seek to find the sample  
size, $n$, such that, for a given $p$ and acceptance precision level $\delta$, the expected length of the confidence interval, $\mathrm{EL}(n, p)\coloneqq \text{E}\left[\Delta(X)\right]$ is equal to $2\delta$, where $\Delta(X)$ is the length of confidence interval, and the expectation is taken over the binomial distribution of $X$.  
The modified Jefferey and the Wilson score methods are described below. 
\begin{enumerate}
\item The Wilson score test confidence interval takes the form
$$\frac{2 X+z_{1-\alpha / 2}^{2}\pm z_{1-\alpha / 2} \sqrt{z_{1-\alpha / 2}^{2}+4 X(1-X / n)}}{2\left(n+z_{1-\alpha / 2}^{2}\right)}$$
This method is derived from Pearson's chi-square test, where the center of the interval is a weighted average of sample proportion and 1/2, such that it is more suitable than the commonly used Wald method for extreme probability or small sample sizes. 
The Wilson method also has the advantage of yielding an analytical
formula for the sample size as follows 
\begin{eqnarray}\label{eqn:wilson}
n_{W}=\frac{-z_{1-\alpha / 2}^{2}[4\delta^{2}-2 p(1-p)]+z_{1-\alpha / 2}^{2} \sqrt{[4\delta^{2}-2 p(1-p)]^{2}-4\delta^{2}\left(4\delta^{2}-1\right)}}{4\delta^{2}},\end{eqnarray}
where $n_{W}$ is the required sample size for a given estimate of $p$ and an acceptance precision level $\delta$.
\item The modified Jeffreys method is derived from a Bayesian approach, which uses the non-informative Jeffrey's prior $\operatorname{Beta}(1/2,1/2)$ to derive the posterior credible interval for $p$, while modifying the formula at the boundary values.
For $1<X<n$, the credible interval is 
$$\left[\operatorname{Beta}_{\alpha / 2}(X+1/2, n-X+1/2), \operatorname{Beta}_{1-\alpha / 2}(X+1/2, n-X+1/2)\right]$$
The expressions when $X$ takes boundary values are provided in Table 1 of \cite{gonccalves2012sample}.
The modified Jeffreys method enjoys similar coverage properties as those of the Wilson score method.  But it has an additional advantage of yielding a credible interval that is equal-tailed. 
For modified Jeffreys,
$$\text{EL}(n; p)= \sum_{X=1}^{n}\Delta(X) {n\choose X} p^{X}(1-p)^{n-X},$$ which is a function of sample size $n$ depending on a given $p$. The sample size can be calculated by solving 
$\text{EL}(n; p)= 2\delta$. It follows that the required sample size using the modified Jeffreys method, denoted as $n_J$,  takes the form
\begin{eqnarray}\label{eqn:jeffrey}
n_{J}=\text{EL}^{-1}(2\delta;p).
\end{eqnarray}
$n_{J}$ does not have a closed form and has to be solved numerically. In practice, it is often calculated by an approximated solution such that $|\text{EL}(n; p)-2\delta|$ is less than a certain tolerance. 
\end{enumerate}

Spatial sampling design involves the determination of sample size, as well as the locations of sampling points. One simple and commonly used spatial sampling design is the uniform random
sampling, where  each location is chosen independently and uniformly within each cluster. However, two complications arise when applying this method for the Arsenic study. First, the number of all available candidate wells are not uniformly distributed in space. Second, a large number of wells have been tested where the sampling locations were arbitrarily chosen before the formal statistical sampling design, which results in a highly unbalanced sampling in space with some areas over-sampled and the other areas insufficiently-sampled. The design for the new sample well locations needs to exclude those previously tested wells. 
Our goal is to sample the candidate wells with the expectation that the combined new sample wells and the previously tested wells are spatially uniformly distributed in each cluster except for the over-sampled areas. 
To achieve this goal, we utilize the connection between the uniform distribution in space and the spatial Poisson point process model, and adopt the thinning sampling idea from the latter.  As a preliminary, we introduce the intensity function of the spatial point processes \citep{diggle1985kernel}, which characterizes the probability that a point occurs in an infinitesimal ball around a given location.  If there is a point process $\mathcal{X}$ on $D\subset\mathbb{R}^2$, let $N(B)$ denote the expected number of points within any subset $B\subset D$.  The intensity function $\lambda(\bfs)$ at location $\bfs\in D$ is defined as,
$$\lambda(\bfs)=\lim_{|b(\bfs)|\rightarrow 0}\frac{N(b(\bfs))}{|b(\bfs)\cap D|}$$
where $b(\bfs)$ denotes a small ball containing $\bfs$, and measure $|\cdot|$ denotes the area.  If $\lambda(\bfs)=\lambda$ is a constant for all $\bfs\in B$, then $\mathcal{X}$ is called a homogeneous point process on $B$, implying the point has the same probability to occur at each location in $B$. Besides, the intensity function determines the expected number of points on $B$ by $\text{E}[N(B)]=\int_{B}\lambda(\bfs)d\bfs$. It is known that, conditional on the number of points, the locations from a homogeneous Poisson point process are uniformly distributed on $B$. 
Therefore, the desired sample well locations have the intensity function $\hat{\lambda}(\bfs)=n_i/a_i$ for $\bfs$ located in cluster $i$, to render the sampled wells evenly-distributed. Here $n_i$ and $a_i$ denote the number of required samples and the area in cluster $i$ respectively.

The detailed sampling algorithm is described below. First, we use the nonparametric intensity estimation approach via $\texttt{R}$ function $\texttt{density.ppp}$ in package $\texttt{spatstat}$ to estimate the candidate well intensity function, denoted as $\hat{\lambda}^{candi}(\bfs)$,
and the previously tested well intensity, denoted $\hat{\lambda}^{exist}(\bfs)$. 
To exclude the previously tested wells in Iowa from new samples, we calculate the target intensity from $\hat{\lambda}^{targ}(\bfs)=\max\{\hat{\lambda}(\bfs)-\hat{\lambda}^{exist}(\bfs),0\}$.
Locations that have negative $\hat{\lambda}(\bfs)-\hat{\lambda}^{exist}(\bfs)$ values correspond to the over-sampled areas where the intensity of previously tested wells exceeds the required sampling density. We will leave them out when drawing new samples. 
Finally, for other areas, each candidate well will be selected with probability $\hat{\lambda}^{targ}(\bfs)/\hat{\lambda}^{candi}(\bfs)$, where $\bfs$ is the location of the candidate well. The last step is based on the assumption that $\hat{\lambda}^{candi}(\bfs)$ is large enough to bound $\hat{\lambda}^{targ}(\bfs)$, and indeed there are adequate wells available in Iowa to meet this assumption. As a result, the algorithm guarantees that the combined new samples and existing samples other than the over-sampled areas will be (nearly) uniformly distributed, and the expected sample size meets the requirement in Table \ref{table:samplesize}.

\section{Results}
\subsection{Descriptive Statistical Analysis Results}
\begin{figure}
\label{fig1}
\centering
\includegraphics[width=5in]{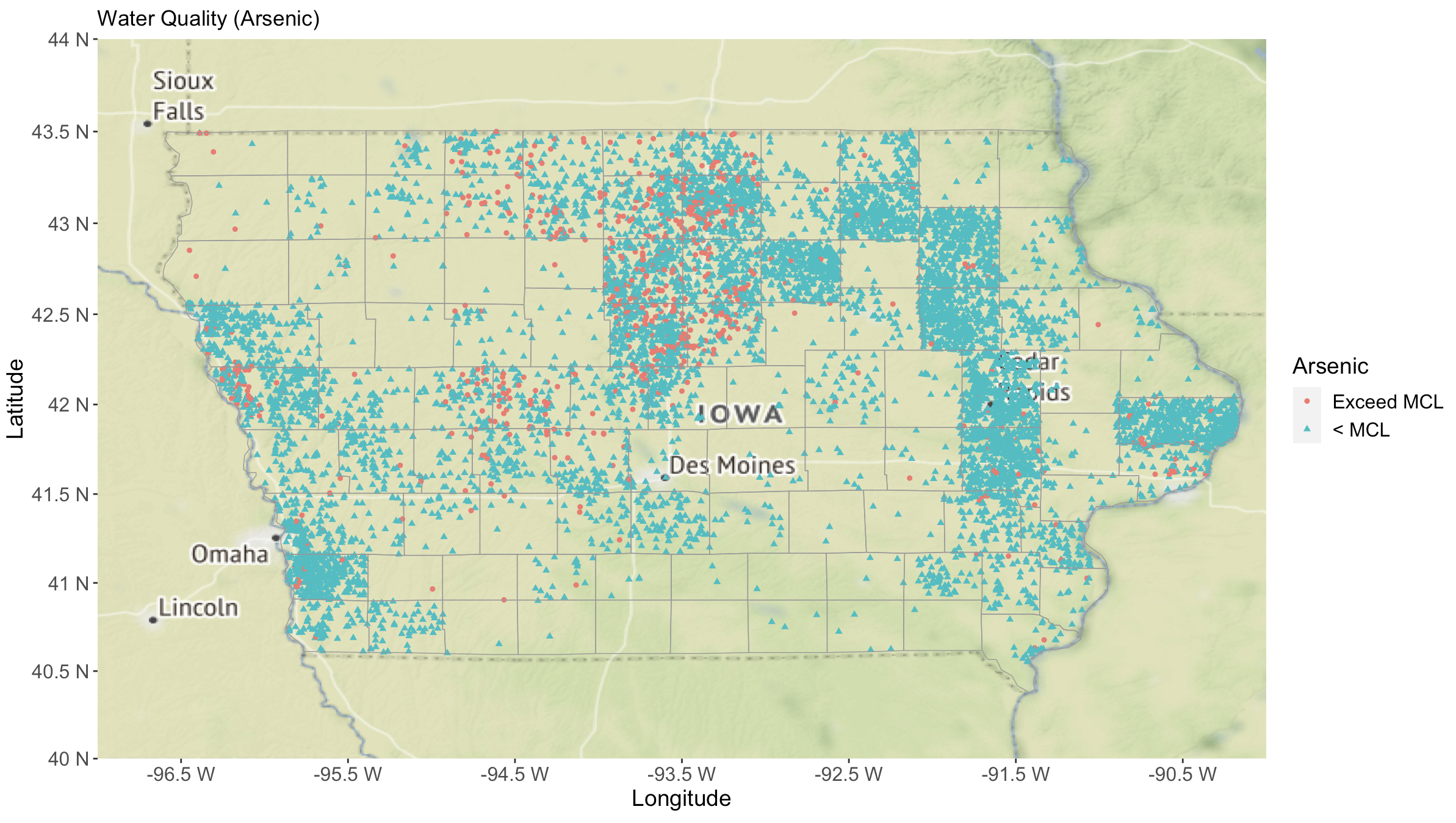}
\caption{Spatial distribution of the Arsenic contamination presence/absence  observations.}
\label{ov1}
\end{figure}

\begin{figure*}
\centering
\includegraphics[width=5in]{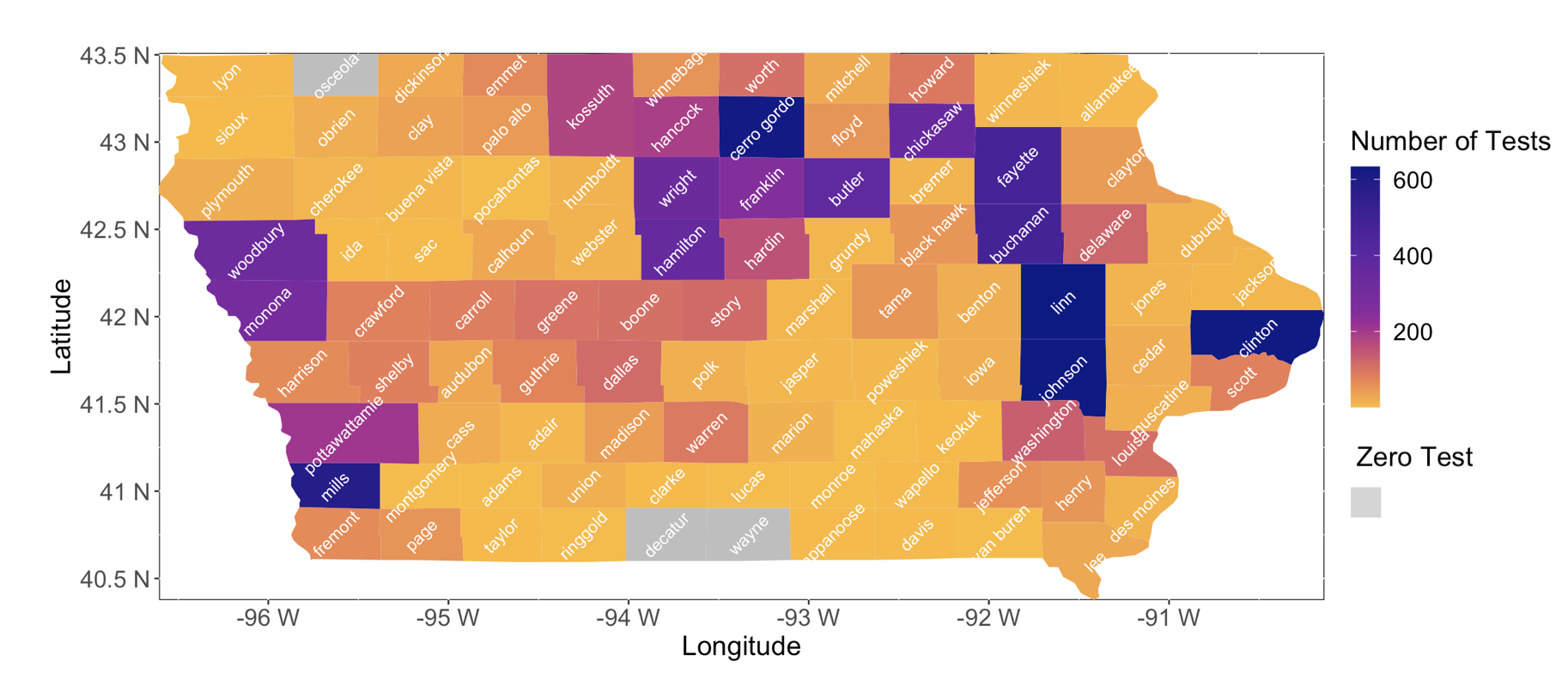}
\caption{The number of tested wells in each county in Iowa.}
\label{dp2}
\end{figure*}

\begin{figure*}
\label{fig3}
\centering
\includegraphics[width=5in]{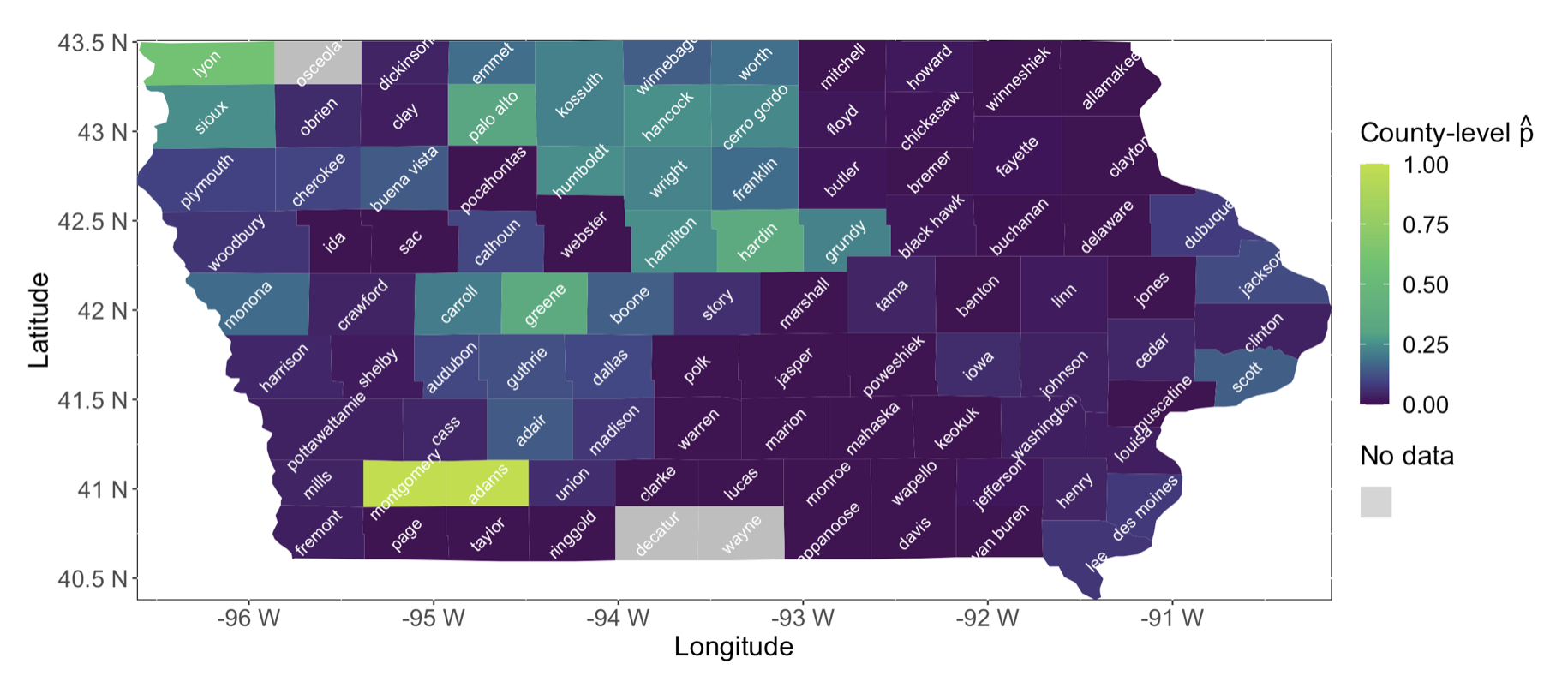}
\caption{This figure illustrates the sample proportion of the contaminated wells among all the observed tested wells for each county in Iowa; Grey color indicates there is no observed data in the county.}
\label{DP3}
\end{figure*}

After the data pre-processing steps, there remain $9842$ observations at different locations. Figure~\ref{ov1} shows the spatial distribution of the observations, and Figure~\ref{dp2} shows the spatial map of the number of observations in each county. 
From the existing tested data, the most tested regions include northern central Iowa, a few counties in the western central, southwestern, and eastern central Iowa regions (Figure~\ref{dp2}). Less than 20\% of the counties have more than 100 tests per county. 
There are fewer tests per county in the southern, northeastern, and northwestern regions. We show in Figure~\ref{DP3} the sample proportion of the contaminated wells among all the tested wells at each county, as a means to visualize a rough empirical estimate of the arsenic risk and its spatial pattern.
Even though we see an uneven testing distribution, which means uneven sampling at the current testing scale, we observe that the arsenic risk characterization appears to be independent of the testing density (Figures~\ref{dp2} and \ref{DP3}). 

\subsection{Risk clusters and regional management}\label{sec:resclust}

\begin{figure*}
\centering
\includegraphics[width=5in]{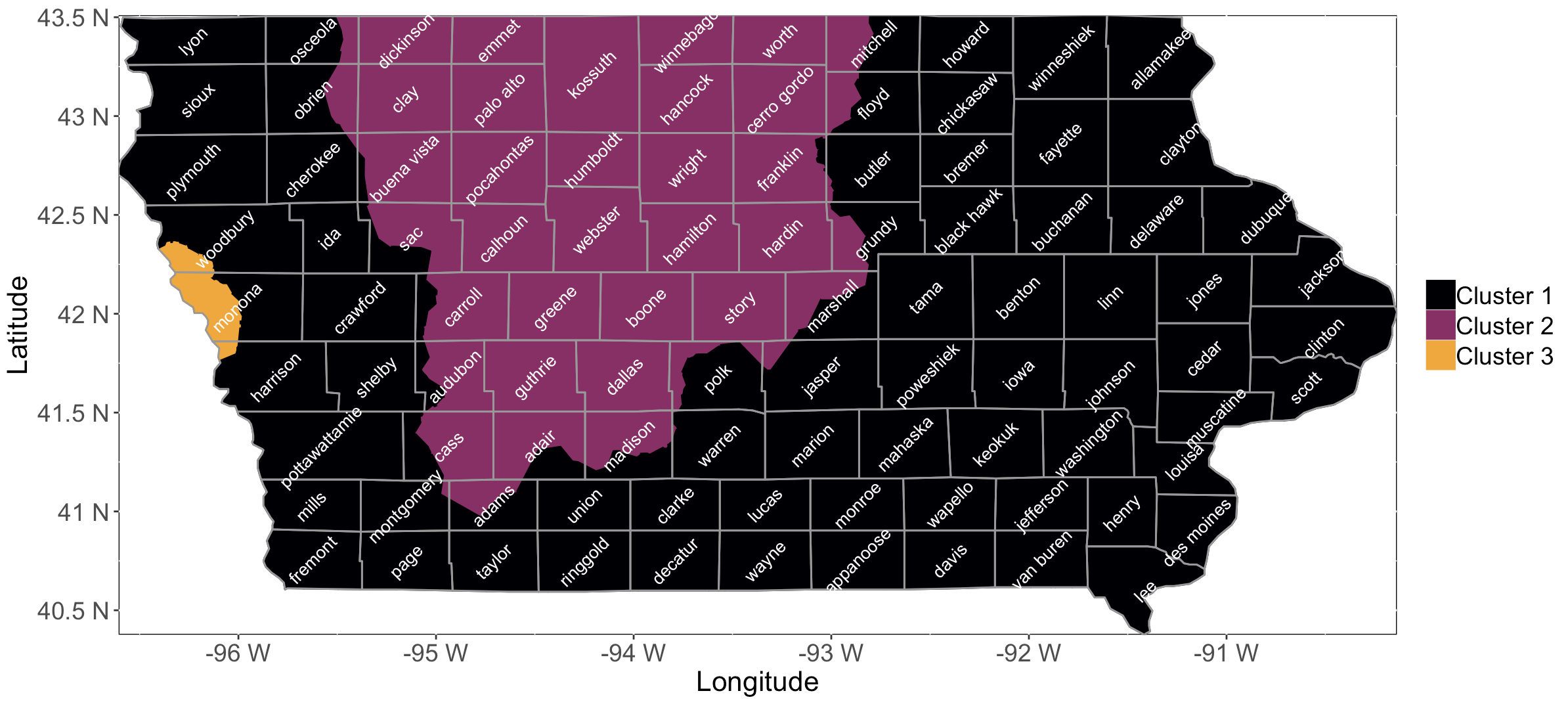}
\caption{Partition of the map in terms of estimated $p$; In cluster 1, $\hat{p}=0.02869485$; In cluster 2, $\hat{p}=0.2088291$; In cluster 3, $\hat{p}=0.3373494$; The numbers of observations in each cluster are $6482$, $3194$ and $166$ respectively.}
\label{u3}
\end{figure*}

Ayotte et al. \cite{ayotte2017estimating} use a predictive logistic regression model to estimate arsenic presence in regions with limited arsenic data. In that study, a total of 20450 domestic well samples are used to develop the model to estimate for the whole conterminous US. Unique to our research, we do not aim to establish a predictive model to accurately predict the arsenic contamination in a given region, as the risk of As has been already recognized by the state and many local health risk management agencies. We aim to utilize the locally clustered arsenic risks to estimate a sample size with minimum bias, which can be managed with appropriately allocated resources. In order to do that, we define the binary existence of arsenic in a given private well is higher than 0.01 mg/L, which is the current EPA regulation level. In other words, we regard if the private well contains less than 0.01 mg/L arsenic, then the health risks are absent in a risk-based sampling scheme.  
We first model and estimate the underlying contamination risk as a spatially clustered function following the method described in Section~\ref{sec:method2} for the straightforward interpretation of the result and easy implementation of the sampling design. The optimization result partitions the whole state into three risk clusters based on the estimated arsenic presence probability (Figure~\ref{u3}). The three risk probabilities (p) are 0.03, 0.21, and 0.33 for clusters 1, 2, and 3, respectively. The risk cluster assignment is consistent with some previous observations and predictions. For example, cluster 1 is largely consistent with the estimations in \cite{ayotte2017estimating}. Cluster 2 is also highlighted with potential high As contamination in the same study. Furthermore, a targeted As study performed in Cerro Gordo County (Northern Central Iowa) has sampled 68 wells over three years \cite{schnoebelen2017elevated}. The study reveals one potential mechanism of As mobilization in the shallow aquifer. The naturally occurring sulfide minerals (typically pyrite) in the bedrock aquifers could be the source of As. Under the oxidizing condition, the As mobilization could happen from rocks to the water. Significantly, the Cerro Gordo study has resulted in a policy change for arsenic testing and well completion locally. Interestingly, cluster 3 at the border of Iowa and Nebraska is identified as a new As "hotspot" in this current study. Notably, the cluster 3 region overlaps with the Missouri alluvial plain. The Missouri River valley contains up to around 150 feet of highly-permeable alluvial sediments \cite{url}. Alluvial sediments could be quite heterogeneous in their gravel, sand, silt, and clay compositions, dependent on the locations. At the same time, those sediments could contain a large percent of argillaceous materials composed of organics, clays, and silts. The presence of argillaceous materials could assist in disseminating arsenic pyrite from the materials themselves or from ferrous hydroxides coating the sand grains, which often contain arsenic as well. Furthermore, diverse geochemical and bacterially mitigated reactions (i.e., oxidation, reduction, adsorption, precipitation, methylation, and volatilization) can participate actively in arsenic recycling within alluvial aquifers. As the alluvial aquifers are largely unconfined, the water table's movement up and down in the aquifer can mobilize arsenic from the argillaceous material or the ferrous hydroxide coating the sand grains through oxidation reactions. The potential high arsenic concentration in the private well in the alluvial plain (i.e., cluster 3) could be attributed to the permeable alluvial sediments and their unique properties.


\subsection{Sample design}

\begin{figure*}
\centering
\includegraphics[width=5in]{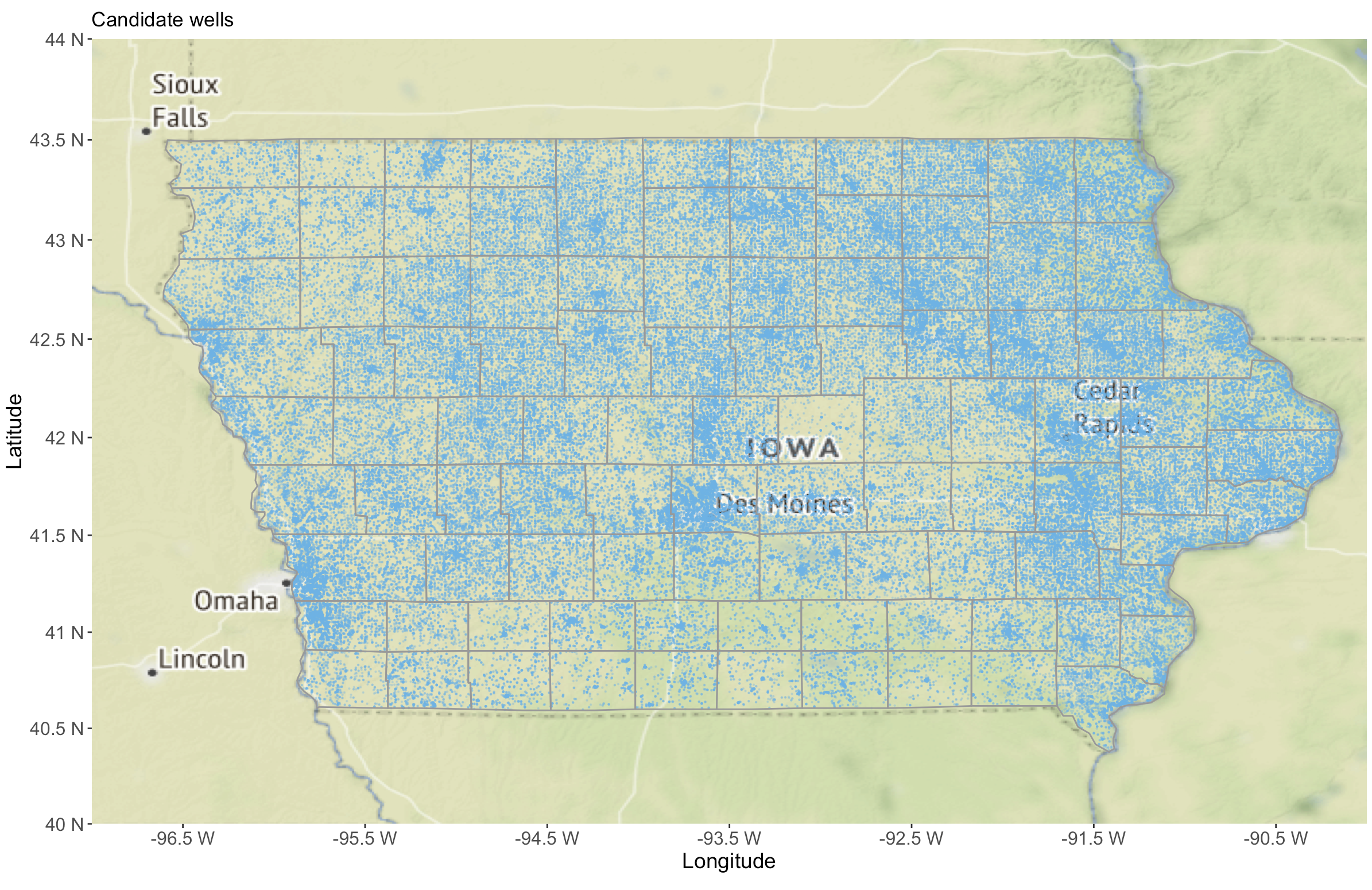}
\caption{Locations of the $291,882$ candidate wells in Iowa, after discarding the wells in absence of their location information.}
\label{c1}
\end{figure*}

Based on the estimated probability clusters, we further estimate the ideal sample size based on various acceptance precision and confidence levels. 
Based on the publicly available database (Iowa Private Well Tracking System), it is estimated there are more than 300,000 private wells in Iowa. Among them, $291,882$ wells are geo-coded. The total geo-coded well population locations are shown in Figure~\ref{c1}, clearly indicating an uneven spatial distribution in Iowa. 
Based on the regional cluster risk probability, we thus define three different cluster regions (clusters 1, 2, and 3) with different risk cluster ranks. For clusters with a reasonable testing coverage, we have three probabilities. For cluster 1,  the estimated probability for As concentration higher than 0.01 mg/L probability is 0.03. For clusters 2 and 3, the probabilities are 0.21 and 0.34, respectively. If we define the precision acceptance as 10\% of the probability, the precision acceptance is 0.003 for cluster 1 ( e.g. 10\% of 0.03), 0.021 for cluster 2, and 0.034 for cluster 3. Table \ref{table:samplesize} provides the calculated required sample size for each cluster under three different confidence levels (90\%, 95\%, and 99\%) using both the Wilson in \eqref{eqn:wilson} and Jeffrey methods in \eqref{eqn:wilson}. For example, at the 95\% confidence interval, the estimated sample size based on the Jeffrey method is 12446 for cluster 1. Applying the same criteria to clusters 2 and 3, the estimated sample size would be 1442 for cluster 2 and 743 for cluster 3. The sample sizes calculated by the Wilson method only slightly differ from those of the Jeffrey method. Accordingly, at the 99\% confidence interval, we estimate that 21456, 2492, and 1282 samples are needed for clusters 1, 2, and 3, respectively. 
 
\begin{table}[width=.9\linewidth,cols=4,pos=h]
\caption{Expected number of well sampling in each cluster;}\label{table:samplesize}
\begin{tabular*}{\tblwidth}{@{} LLLLL@{} }
\toprule
Confidence Level & Method & Cluster 1 & Cluster 2 & Cluster 3 \\
\midrule
$90\%$ & Wilson & 8766 & 1017 & 523\\
& Jeffrey & 8746 & 1015 & 523 \\
$95\%$ & Wilson & 12446 & 1444 & 743 \\
 & Jeffrey & 12420 & 1442 & 743 \\
$99\%$ & Wilson & 21497 & 2493 & 1282 \\
& Jeffrey & 21456 & 2492 & 1284\\
\bottomrule
\end{tabular*}
\end{table}

\begin{figure*}
\centering
\includegraphics[width=5in]{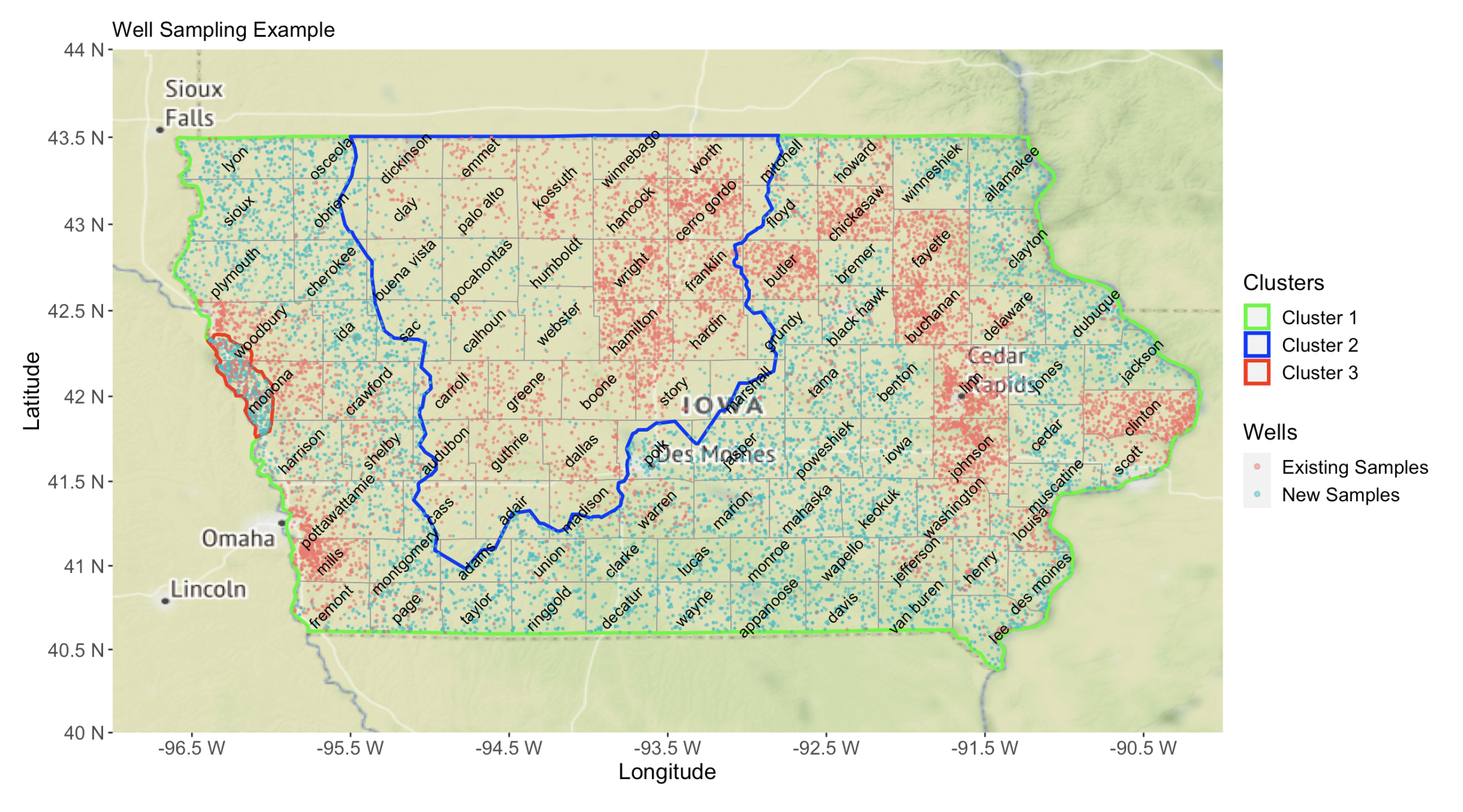}
\caption{An example of sampling results; $8174$, $313$ and $586$ additional wells are sampled in each cluster respectively in this example.}
\label{sampling3}
\end{figure*}

In the existing As data set, there are 6482, 3194, and 166 tested wells already collected from clusters 1, 2 , and 3, respectively. It is noted that the sample size of the tested wells within each cluster constitutes a large proportion or exceeds the required sample size calculated in Table \ref{table:samplesize}. However, we recognize the current As data collection is operated at the county level since the local environmental health jurisdiction resides in each county.  
County level data generation results in an uneven spatial distribution of sampling locations for the whole state discussed in Section \ref{sec:resclust}. Therefore, although some areas are over-sampled, new samples still need to be collected at those places that are only sparsely sampled previously. 

We follow the method presented in Section \ref{sec:sample} to determine the locations of new sampling locations. We use the private wells in the current Iowa PWTS database as the target sampling population (Figure \ref{c1}).  The goal is to achieve a spatially balanced sampling design that meets the required sample size, while accounting for the fact that both the candidate wells and existing tested wells are distributed highly non-uniformly in space. To illustrate, we give an example of the sampling scenario using the sample size calculated from the Wilson method for the $95\%$ CI in Figure \ref{sampling3}.  
The dense red point clouds reveal the previously over sampled areas in this Figure.   
Cluster 2 has the largest proportion of previously over-sampled areas. Only a relatively small number of additional wells (marked by green dots) need to be sampled, mostly are located in the middle west of this region. In contrast, most areas in cluster 1 have not been sampled and tested previously, with exceptions in several counties (e.g., Buchanan, Butler, and Clinton). In cluster 3, although the spatial coverage of the existing tested samples is nearly uniform, our method suggests that an additional number of wells need to be collected to achieve the desired confidence level and precision accuracy. 
Overall, it is noted that the locations of samples in Figure \ref{sampling3} appear to be uniformly distributed except for the previously over-sampled areas. Looking more closely, we observe that the intensity/density of samples differs across the identified risk clusters, due to adopting a spatially adaptive sampling design according to each cluster's own contamination risk.

\section{Discussions}

It is commonly recognized that many conditions such as geological, geochemical, and hydrologic variables, impact arsenic presence in groundwater. For example, It has been observed high arsenic concentrations are often found in more arid western US \citep{ayotte2017estimating}. Furthermore, precipitation and recharge show significant correlations with arsenic concentrations in domestic wells in the conterminous US. Among various conditions, glaciated terrain, bedrock geology, soil hydrology, soil tile drainage, water table depth and climate factors can also impact arsenic concentrations in groundwater. Particularly, Iowa's groundwater resources are majorly surficial aquifers and bedrock aquifers. For a long history contacting with glaciers, many parts of Iowa soil/dirt contain glacier age materials with moderate to low permeability. The water table beneath those materials occurs at relatively shallow depths and varies from 3 to 30 feet below ground \citep{prior2003iowa}.The micro-environment such as pH, soil, and water bacterial activity, oxidation and reduction reactions (Redox), coexistence with other elements (e.g., iron) can also play a significant role in arsenic concentration in groundwater. Taking account of all those macro and micro-environmental conditions is a shared challenge for all current available predictive models to estimate arsenic concentrations at the county, state/province, or region levels.    

There are several potential benefits to adopt the proposed sampling design. First, the sample size estimate suggests future feasible random sampling targets, given the total Iowa private well population. As the sample sizes are dependent on the arsenic probabilities, we present options for the same probability with different sampling precision goals. We also recognize there are regions with too few or no data points (Figure~\ref{DP3}, thus warrant further sampling for probability estimate). Second, the method developed in this study helps pinpoint future sampling locations with adequate statistical power. From the resource management perspective, future planning can prioritize the high-risk well sampling, eliminate redundant testing, and collect representative samples for risk assessment purposes. In practice, sample collections and management  are often conducted at certain administration levels. It is desired to develop a sampling design method that is easy and fast to implement at each administration unit. Third, this design presents future opportunities to investigate practical solutions to coordinate joint efforts across counties for the efficient implementation of the sampling design method. 

Moving forward, this work could be further refined in several ways. First, the estimator we obtained by optimizing the regularized log-likelihood function does not come with  an uncertainty measure. As such, the sample size calculation is only based on a point estimate of the contamination risk. A potential solution is to consider a Bayesian version of the method. In principle, the modified Jeffrey's method for sample size calculation can be adapted to account for the uncertainty in the estimate of the contamination probability $p$, where the expected length of the confidence interval used in \eqref{eqn:jeffrey} can be taken over both the distributions of $p$ and the binomial random variable $X$ instead of $X$ only. Second, we assume that the clusters of wells are spatially contiguous, and the contiguity constraint is defined with respect to the choice of a spatial graph. However, in practice, the spatial contiguity constraint may not dominate the clustering configuration globally, in the sense that two or more locally contiguous clusters that are remote in space may actually have very similar arsenic concentrations, and hence should be classified into the same cluster. The method presented in this paper needs to be modified to handle the case with both globally dis-contiguous and locally contiguous clusters. One idea is to perform a two-step analysis, where in the first step we obtain local spatial clusters from the method presented in this paper, and in the second step, conduct another clustering analysis without any spatial constraint based on the local clustering results from the first step. Third, the model can be further improved with more representative samples. As we noted, there are counties without testing data, which presents a gap for risk analysis. We expect collecting data in those regions helps build a more comprehensive evaluation of arsenic health risk at the state level. Overall, the current study presents a targeted approach to save cost and time for effective public health management strategy. The rational sampling design focuses on risk categories, which assures that preventive measures and mitigation practices are implemented where most needed.     

\bibliographystyle{model1-num-names}
\bibliography{Bibliography-MM-MC}

\begin{thebibliography}{45}
\expandafter\ifx\csname natexlab\endcsname\relax\def\natexlab#1{#1}\fi
\providecommand{\url}[1]{\texttt{#1}}
\providecommand{\href}[2]{#2}
\providecommand{\path}[1]{#1}
\providecommand{\DOIprefix}{doi:}
\providecommand{\ArXivprefix}{arXiv:}
\providecommand{\URLprefix}{URL: }
\providecommand{\Pubmedprefix}{pmid:}
\providecommand{\doi}[1]{\href{http://dx.doi.org/#1}{\path{#1}}}
\providecommand{\Pubmed}[1]{\href{pmid:#1}{\path{#1}}}
\providecommand{\bibinfo}[2]{#2}
\ifx\xfnm\relax \def\xfnm[#1]{\unskip,\space#1}\fi
\bibitem[{Podgorski and Berg(2020)}]{podgorski2020global}
\bibinfo{author}{J.~Podgorski}, \bibinfo{author}{M.~Berg},
\newblock \bibinfo{title}{Global threat of arsenic in groundwater},
\newblock \bibinfo{journal}{Science} \bibinfo{volume}{368}
  (\bibinfo{year}{2020}) \bibinfo{pages}{845--850}.
\bibitem[{DeSimone and Hamilton(2009)}]{desimone2009quality}
\bibinfo{author}{L.~A. DeSimone}, \bibinfo{author}{P.~A. Hamilton},
  \bibinfo{title}{{Quality of water from domestic wells in principal aquifers
  of the United States, 1991-2004}}, \bibinfo{publisher}{US Department of the
  Interior, US Geological Survey}, \bibinfo{year}{2009}.
\bibitem[{Almberg et~al.(2017)Almberg, Turyk, Jones, Rankin, Freels, Graber,
  and Stayner}]{almberg2017arsenic}
\bibinfo{author}{K.~S. Almberg}, \bibinfo{author}{M.~E. Turyk},
  \bibinfo{author}{R.~M. Jones}, \bibinfo{author}{K.~Rankin},
  \bibinfo{author}{S.~Freels}, \bibinfo{author}{J.~M. Graber},
  \bibinfo{author}{L.~T. Stayner},
\newblock \bibinfo{title}{{Arsenic in drinking water and adverse birth outcomes
  in Ohio}},
\newblock \bibinfo{journal}{Environmental research} \bibinfo{volume}{157}
  (\bibinfo{year}{2017}) \bibinfo{pages}{52--59}.
\bibitem[{Vahter(2009)}]{vahter2009effects}
\bibinfo{author}{M.~Vahter},
\newblock \bibinfo{title}{Effects of arsenic on maternal and fetal health},
\newblock \bibinfo{journal}{Annual review of nutrition} \bibinfo{volume}{29}
  (\bibinfo{year}{2009}) \bibinfo{pages}{381--399}.
\bibitem[{Sohel et~al.(2009)Sohel, Persson, Rahman, Streatfield, Yunus,
  Ekstr{\"o}m, and Vahter}]{sohel2009arsenic}
\bibinfo{author}{N.~Sohel}, \bibinfo{author}{L.~{\AA}. Persson},
  \bibinfo{author}{M.~Rahman}, \bibinfo{author}{P.~K. Streatfield},
  \bibinfo{author}{M.~Yunus}, \bibinfo{author}{E.-C. Ekstr{\"o}m},
  \bibinfo{author}{M.~Vahter},
\newblock \bibinfo{title}{{Arsenic in drinking water and adult mortality: a
  population-based cohort study in rural Bangladesh}},
\newblock \bibinfo{journal}{Epidemiology}  (\bibinfo{year}{2009})
  \bibinfo{pages}{824--830}.
\bibitem[{Argos et~al.(2012)Argos, Ahsan, and Graziano}]{argos2012arsenic}
\bibinfo{author}{M.~Argos}, \bibinfo{author}{H.~Ahsan}, \bibinfo{author}{J.~H.
  Graziano},
\newblock \bibinfo{title}{Arsenic and human health: epidemiologic progress and
  public health implications},
\newblock \bibinfo{journal}{Reviews on environmental health}
  \bibinfo{volume}{27} (\bibinfo{year}{2012}) \bibinfo{pages}{191--195}.
\bibitem[{Bloom et~al.(2014)Bloom, Surdu, Neamtiu, and
  Gurzau}]{bloom2014maternal}
\bibinfo{author}{M.~S. Bloom}, \bibinfo{author}{S.~Surdu},
  \bibinfo{author}{I.~A. Neamtiu}, \bibinfo{author}{E.~S. Gurzau},
\newblock \bibinfo{title}{Maternal arsenic exposure and birth outcomes: a
  comprehensive review of the epidemiologic literature focused on drinking
  water},
\newblock \bibinfo{journal}{International journal of hygiene and environmental
  health} \bibinfo{volume}{217} (\bibinfo{year}{2014})
  \bibinfo{pages}{709--719}.
\bibitem[{Association(2020)}]{national2016groundwater}
\bibinfo{author}{N.~G. Association}, \bibinfo{title}{{Groundwater use in the
  United States of America}}, \bibinfo{year}{2020}. \URLprefix
  \url{https://www.ngwa.org/docs/default-source/default-document-library/groundwater/usa-groundwater-use-fact-sheet.pdf?sfvrsn=5c7a0db8_4}.
\bibitem[{Schnoebelen et~al.(2017)Schnoebelen, Walsh, Hanft, Hernandez-Murcia,
  and Fields}]{schnoebelen2017elevated}
\bibinfo{author}{D.~J. Schnoebelen}, \bibinfo{author}{S.~Walsh},
  \bibinfo{author}{B.~Hanft}, \bibinfo{author}{O.~E. Hernandez-Murcia},
  \bibinfo{author}{C.~Fields},
\newblock \bibinfo{title}{{Elevated Arsenic in Private Wells of Cerro Gordo
  County, Iowa: Causes and Policy Changes.}},
\newblock \bibinfo{journal}{Journal of Environmental Health}
  \bibinfo{volume}{79} (\bibinfo{year}{2017}).
\bibitem[{Minkkinen(2004)}]{minkkinen2004practical}
\bibinfo{author}{P.~Minkkinen},
\newblock \bibinfo{title}{Practical applications of sampling theory},
\newblock \bibinfo{journal}{Chemometrics and intelligent laboratory systems}
  \bibinfo{volume}{74} (\bibinfo{year}{2004}) \bibinfo{pages}{85--94}.
\bibitem[{(USEPA)(2002)}]{us2002guidance}
\bibinfo{author}{U.~E. P.~A. (USEPA)}, \bibinfo{title}{Guidance on choosing a
  sampling design for environmental data collection}, \bibinfo{year}{2002}.
\bibitem[{Gon{\c{c}}alves et~al.(2012)Gon{\c{c}}alves, de~Oliveira, Pascoal,
  and Pires}]{gonccalves2012sample}
\bibinfo{author}{L.~Gon{\c{c}}alves}, \bibinfo{author}{M.~R. de~Oliveira},
  \bibinfo{author}{C.~Pascoal}, \bibinfo{author}{A.~Pires},
\newblock \bibinfo{title}{Sample size for estimating a binomial proportion:
  comparison of different methods},
\newblock \bibinfo{journal}{Journal of Applied Statistics} \bibinfo{volume}{39}
  (\bibinfo{year}{2012}) \bibinfo{pages}{2453--2473}.
\bibitem[{Lee et~al.(2016)Lee, Herrman, and Dai}]{lee2016application}
\bibinfo{author}{K.-M. Lee}, \bibinfo{author}{T.~J. Herrman},
  \bibinfo{author}{S.~Y. Dai},
\newblock \bibinfo{title}{Application and validation of a statistically derived
  risk-based sampling plan to improve efficiency of inspection and
  enforcement},
\newblock \bibinfo{journal}{Food Control} \bibinfo{volume}{64}
  (\bibinfo{year}{2016}) \bibinfo{pages}{135--141}.
\bibitem[{Sep{\'u}lveda and Drakeley(2015)}]{sepulveda2015sample}
\bibinfo{author}{N.~Sep{\'u}lveda}, \bibinfo{author}{C.~Drakeley},
\newblock \bibinfo{title}{Sample size determination for estimating antibody
  seroconversion rate under stable malaria transmission intensity},
\newblock \bibinfo{journal}{Malaria journal} \bibinfo{volume}{14}
  (\bibinfo{year}{2015}) \bibinfo{pages}{141}.
\bibitem[{Joseph and Reinhold(2005)}]{joseph2005statistical}
\bibinfo{author}{L.~Joseph}, \bibinfo{author}{C.~Reinhold},
\newblock \bibinfo{title}{Statistical inference for continuous variables},
\newblock \bibinfo{journal}{American Journal of Roentgenology}
  \bibinfo{volume}{184} (\bibinfo{year}{2005}) \bibinfo{pages}{1047--1056}.
\bibitem[{Amini et~al.(2008)Amini, Abbaspour, Berg, Winkel, Hug, Hoehn, Yang,
  and Johnson}]{amini2008statistical}
\bibinfo{author}{M.~Amini}, \bibinfo{author}{K.~C. Abbaspour},
  \bibinfo{author}{M.~Berg}, \bibinfo{author}{L.~Winkel},
  \bibinfo{author}{S.~J. Hug}, \bibinfo{author}{E.~Hoehn},
  \bibinfo{author}{H.~Yang}, \bibinfo{author}{C.~A. Johnson},
\newblock \bibinfo{title}{Statistical modeling of global geogenic arsenic
  contamination in groundwater},
\newblock \bibinfo{journal}{Environmental science \& technology}
  \bibinfo{volume}{42} (\bibinfo{year}{2008}) \bibinfo{pages}{3669--3675}.
\bibitem[{Ayotte et~al.(2006)Ayotte, Nolan, Nuckols, Cantor, Robinson, Baris,
  Hayes, Karagas, Bress, Silverman et~al.}]{ayotte2006modeling}
\bibinfo{author}{J.~D. Ayotte}, \bibinfo{author}{B.~T. Nolan},
  \bibinfo{author}{J.~R. Nuckols}, \bibinfo{author}{K.~P. Cantor},
  \bibinfo{author}{G.~R. Robinson}, \bibinfo{author}{D.~Baris},
  \bibinfo{author}{L.~Hayes}, \bibinfo{author}{M.~Karagas},
  \bibinfo{author}{W.~Bress}, \bibinfo{author}{D.~T. Silverman}, et~al.,
\newblock \bibinfo{title}{{Modeling the probability of arsenic in groundwater
  in New England as a tool for exposure assessment}},
\newblock \bibinfo{journal}{Environmental science \& technology}
  \bibinfo{volume}{40} (\bibinfo{year}{2006}) \bibinfo{pages}{3578--3585}.
\bibitem[{Winkel et~al.(2008)Winkel, Berg, Amini, Hug, and
  Johnson}]{winkel2008predicting}
\bibinfo{author}{L.~Winkel}, \bibinfo{author}{M.~Berg},
  \bibinfo{author}{M.~Amini}, \bibinfo{author}{S.~J. Hug},
  \bibinfo{author}{C.~A. Johnson},
\newblock \bibinfo{title}{{Predicting groundwater arsenic contamination in
  Southeast Asia from surface parameters}},
\newblock \bibinfo{journal}{Nature Geoscience} \bibinfo{volume}{1}
  (\bibinfo{year}{2008}) \bibinfo{pages}{536--542}.
\bibitem[{Podgorski et~al.(2017)Podgorski, Eqani, Khanam, Ullah, Shen, and
  Berg}]{podgorski2017extensive}
\bibinfo{author}{J.~E. Podgorski}, \bibinfo{author}{S.~A. M. A.~S. Eqani},
  \bibinfo{author}{T.~Khanam}, \bibinfo{author}{R.~Ullah},
  \bibinfo{author}{H.~Shen}, \bibinfo{author}{M.~Berg},
\newblock \bibinfo{title}{{Extensive arsenic contamination in high-pH
  unconfined aquifers in the Indus Valley}},
\newblock \bibinfo{journal}{Science advances} \bibinfo{volume}{3}
  (\bibinfo{year}{2017}) \bibinfo{pages}{e1700935}.
\bibitem[{Winkel et~al.(2011)Winkel, Trang, Lan, Stengel, Amini, Ha, Viet, and
  Berg}]{winkel2011arsenic}
\bibinfo{author}{L.~H. Winkel}, \bibinfo{author}{P.~T.~K. Trang},
  \bibinfo{author}{V.~M. Lan}, \bibinfo{author}{C.~Stengel},
  \bibinfo{author}{M.~Amini}, \bibinfo{author}{N.~T. Ha},
  \bibinfo{author}{P.~H. Viet}, \bibinfo{author}{M.~Berg},
\newblock \bibinfo{title}{Arsenic pollution of groundwater in vietnam
  exacerbated by deep aquifer exploitation for more than a century},
\newblock \bibinfo{journal}{Proceedings of the National Academy of Sciences}
  \bibinfo{volume}{108} (\bibinfo{year}{2011}) \bibinfo{pages}{1246--1251}.
\bibitem[{Rodr{\'\i}guez-Lado et~al.(2013)Rodr{\'\i}guez-Lado, Sun, Berg,
  Zhang, Xue, Zheng, and Johnson}]{rodriguez2013groundwater}
\bibinfo{author}{L.~Rodr{\'\i}guez-Lado}, \bibinfo{author}{G.~Sun},
  \bibinfo{author}{M.~Berg}, \bibinfo{author}{Q.~Zhang},
  \bibinfo{author}{H.~Xue}, \bibinfo{author}{Q.~Zheng}, \bibinfo{author}{C.~A.
  Johnson},
\newblock \bibinfo{title}{{Groundwater arsenic contamination throughout
  China}},
\newblock \bibinfo{journal}{Science} \bibinfo{volume}{341}
  (\bibinfo{year}{2013}) \bibinfo{pages}{866--868}.
\bibitem[{Yang et~al.(2012)Yang, Jung, Marvinney, Culbertson, and
  Zheng}]{yang2012can}
\bibinfo{author}{Q.~Yang}, \bibinfo{author}{H.~B. Jung}, \bibinfo{author}{R.~G.
  Marvinney}, \bibinfo{author}{C.~W. Culbertson}, \bibinfo{author}{Y.~Zheng},
\newblock \bibinfo{title}{Can arsenic occurrence rates in bedrock aquifers be
  predicted?},
\newblock \bibinfo{journal}{Environmental science \& technology}
  \bibinfo{volume}{46} (\bibinfo{year}{2012}) \bibinfo{pages}{2080--2087}.
\bibitem[{Ayotte et~al.(2017)Ayotte, Medalie, Qi, Backer, and
  Nolan}]{ayotte2017estimating}
\bibinfo{author}{J.~D. Ayotte}, \bibinfo{author}{L.~Medalie},
  \bibinfo{author}{S.~L. Qi}, \bibinfo{author}{L.~C. Backer},
  \bibinfo{author}{B.~T. Nolan},
\newblock \bibinfo{title}{{Estimating the high-arsenic domestic-well population
  in the conterminous United States}},
\newblock \bibinfo{journal}{Environmental science \& technology}
  \bibinfo{volume}{51} (\bibinfo{year}{2017}) \bibinfo{pages}{12443--12454}.
\bibitem[{Ayotte et~al.(2016)Ayotte, Nolan, and
  Gronberg}]{ayotte2016predicting}
\bibinfo{author}{J.~D. Ayotte}, \bibinfo{author}{B.~T. Nolan},
  \bibinfo{author}{J.~A. Gronberg},
\newblock \bibinfo{title}{{Predicting arsenic in drinking water wells of the
  Central Valley, California}},
\newblock \bibinfo{journal}{Environmental Science \& Technology}
  \bibinfo{volume}{50} (\bibinfo{year}{2016}) \bibinfo{pages}{7555--7563}.
\bibitem[{Erickson et~al.(2018)Erickson, Elliott, Christenson, and
  Krall}]{erickson2018predicting}
\bibinfo{author}{M.~L. Erickson}, \bibinfo{author}{S.~M. Elliott},
  \bibinfo{author}{C.~Christenson}, \bibinfo{author}{A.~L. Krall},
\newblock \bibinfo{title}{{Predicting geogenic Arsenic in Drinking Water Wells
  in Glacial Aquifers, North-Central USA: Accounting for Depth-Dependent
  Features}},
\newblock \bibinfo{journal}{Water Resources Research} \bibinfo{volume}{54}
  (\bibinfo{year}{2018}) \bibinfo{pages}{10--172}.
\bibitem[{Tan et~al.(2020)Tan, Yang, and Zheng}]{tan2020machine}
\bibinfo{author}{Z.~Tan}, \bibinfo{author}{Q.~Yang},
  \bibinfo{author}{Y.~Zheng},
\newblock \bibinfo{title}{{Machine Learning Models of Groundwater Arsenic
  Spatial Distribution in Bangladesh: Influence of Holocene Sediment
  Depositional History}},
\newblock \bibinfo{journal}{Environmental Science \& Technology}
  \bibinfo{volume}{54} (\bibinfo{year}{2020}) \bibinfo{pages}{9454--9463}.
\bibitem[{Zhu and Stein(2006)}]{zhu2006spatial}
\bibinfo{author}{Z.~Zhu}, \bibinfo{author}{M.~L. Stein},
\newblock \bibinfo{title}{Spatial sampling design for prediction with estimated
  parameters},
\newblock \bibinfo{journal}{Journal of agricultural, biological, and
  environmental statistics} \bibinfo{volume}{11} (\bibinfo{year}{2006})
  \bibinfo{pages}{24}.
\bibitem[{Diggle et~al.(2010)Diggle, Menezes, and
  Su}]{diggle2010geostatistical}
\bibinfo{author}{P.~J. Diggle}, \bibinfo{author}{R.~Menezes},
  \bibinfo{author}{T.-l. Su},
\newblock \bibinfo{title}{Geostatistical inference under preferential
  sampling},
\newblock \bibinfo{journal}{Journal of the Royal Statistical Society: Series C
  (Applied Statistics)} \bibinfo{volume}{59} (\bibinfo{year}{2010})
  \bibinfo{pages}{191--232}.
\bibitem[{Tibshirani et~al.(2005)Tibshirani, Saunders, Rosset, Zhu, and
  Knight}]{tibshirani2005sparsity}
\bibinfo{author}{R.~Tibshirani}, \bibinfo{author}{M.~Saunders},
  \bibinfo{author}{S.~Rosset}, \bibinfo{author}{J.~Zhu},
  \bibinfo{author}{K.~Knight},
\newblock \bibinfo{title}{Sparsity and smoothness via the fused lasso},
\newblock \bibinfo{journal}{Journal of the Royal Statistical Society: Series B
  (Statistical Methodology)} \bibinfo{volume}{67} (\bibinfo{year}{2005})
  \bibinfo{pages}{91--108}.
\bibitem[{of~Public~Health(2016)}]{Iowa}
\bibinfo{author}{I.~D. of~Public~Health}, \bibinfo{title}{{Iowa Administrative
  Code 641, Chapter 24, Private Well Testing, Reconstruction, and Plugging-
  Grants to Counties }}, \bibinfo{year}{2016}.
\bibitem[{url(2021)}]{url2}
\bibinfo{title}{{Arsenic Testing Results in Private Well Water}},
  \bibinfo{howpublished}{Iowa Department of Public Health},
  \bibinfo{year}{accessed January 2021}. \URLprefix
  \url{https://tracking.idph.iowa.gov/Environment/Private-Well-Water}.
\bibitem[{B{\"u}hlmann and Van De~Geer(2011)}]{buhlmann2011statistics}
\bibinfo{author}{P.~B{\"u}hlmann}, \bibinfo{author}{S.~Van De~Geer},
  \bibinfo{title}{Statistics for high-dimensional data: methods, theory and
  applications}, \bibinfo{publisher}{Springer Science \& Business Media},
  \bibinfo{year}{2011}.
\bibitem[{Tibshirani et~al.(2011)Tibshirani, Taylor
  et~al.}]{tibshirani2011solution}
\bibinfo{author}{R.~J. Tibshirani}, \bibinfo{author}{J.~Taylor}, et~al.,
\newblock \bibinfo{title}{The solution path of the generalized lasso},
\newblock \bibinfo{journal}{The Annals of Statistics} \bibinfo{volume}{39}
  (\bibinfo{year}{2011}) \bibinfo{pages}{1335--1371}.
\bibitem[{Li and Sang(2019)}]{li2019spatial}
\bibinfo{author}{F.~Li}, \bibinfo{author}{H.~Sang},
\newblock \bibinfo{title}{Spatial homogeneity pursuit of regression
  coefficients for large datasets},
\newblock \bibinfo{journal}{Journal of the American Statistical Association}
  \bibinfo{volume}{114} (\bibinfo{year}{2019}) \bibinfo{pages}{1050--1062}.
\bibitem[{Madrid~Padilla et~al.(2020)Madrid~Padilla, Sharpnack, Chen, and
  Witten}]{madrid2020adaptive}
\bibinfo{author}{O.~H. Madrid~Padilla}, \bibinfo{author}{J.~Sharpnack},
  \bibinfo{author}{Y.~Chen}, \bibinfo{author}{D.~M. Witten},
\newblock \bibinfo{title}{Adaptive nonparametric regression with the k-nearest
  neighbour fused lasso},
\newblock \bibinfo{journal}{Biometrika} \bibinfo{volume}{107}
  (\bibinfo{year}{2020}) \bibinfo{pages}{293--310}.
\bibitem[{Beck and Teboulle(2009)}]{beck2009fast}
\bibinfo{author}{A.~Beck}, \bibinfo{author}{M.~Teboulle},
\newblock \bibinfo{title}{Fast gradient-based algorithms for constrained total
  variation image denoising and deblurring problems},
\newblock \bibinfo{journal}{IEEE transactions on image processing}
  \bibinfo{volume}{18} (\bibinfo{year}{2009}) \bibinfo{pages}{2419--2434}.
\bibitem[{Boyd et~al.(2011)Boyd, Parikh, Chu, Peleato, Eckstein
  et~al.}]{boyd2011distributed}
\bibinfo{author}{S.~Boyd}, \bibinfo{author}{N.~Parikh},
  \bibinfo{author}{E.~Chu}, \bibinfo{author}{B.~Peleato},
  \bibinfo{author}{J.~Eckstein}, et~al.,
\newblock \bibinfo{title}{Distributed optimization and statistical learning via
  the alternating direction method of multipliers},
\newblock \bibinfo{journal}{Foundations and Trends{\textregistered} in Machine
  learning} \bibinfo{volume}{3} (\bibinfo{year}{2011}) \bibinfo{pages}{1--122}.
\bibitem[{Wahlberg et~al.(2012)Wahlberg, Boyd, Annergren, and
  Wang}]{wahlberg2012admm}
\bibinfo{author}{B.~Wahlberg}, \bibinfo{author}{S.~Boyd},
  \bibinfo{author}{M.~Annergren}, \bibinfo{author}{Y.~Wang},
\newblock \bibinfo{title}{{An ADMM algorithm for a class of total variation
  regularized estimation problems}},
\newblock \bibinfo{journal}{IFAC Proceedings Volumes} \bibinfo{volume}{45}
  (\bibinfo{year}{2012}) \bibinfo{pages}{83--88}.
\bibitem[{Golub et~al.(1979)Golub, Heath, and Wahba}]{golub1979generalized}
\bibinfo{author}{G.~H. Golub}, \bibinfo{author}{M.~Heath},
  \bibinfo{author}{G.~Wahba},
\newblock \bibinfo{title}{Generalized cross-validation as a method for choosing
  a good ridge parameter},
\newblock \bibinfo{journal}{Technometrics} \bibinfo{volume}{21}
  (\bibinfo{year}{1979}) \bibinfo{pages}{215--223}.
\bibitem[{Schwarz(1978)}]{schwarz1978estimating}
\bibinfo{author}{G.~Schwarz},
\newblock \bibinfo{title}{Estimating the dimension of a model},
\newblock \bibinfo{journal}{The annals of statistics} \bibinfo{volume}{6}
  (\bibinfo{year}{1978}) \bibinfo{pages}{461--464}.
\bibitem[{Chen and Chen(2008)}]{chen2008extended}
\bibinfo{author}{J.~Chen}, \bibinfo{author}{Z.~Chen},
\newblock \bibinfo{title}{{Extended Bayesian information criteria for model
  selection with large model spaces}},
\newblock \bibinfo{journal}{Biometrika} \bibinfo{volume}{95}
  (\bibinfo{year}{2008}) \bibinfo{pages}{759--771}.
\bibitem[{Newcombe(1998)}]{newcombe1998two}
\bibinfo{author}{R.~G. Newcombe},
\newblock \bibinfo{title}{Two-sided confidence intervals for the single
  proportion: comparison of seven methods},
\newblock \bibinfo{journal}{Statistics in medicine} \bibinfo{volume}{17}
  (\bibinfo{year}{1998}) \bibinfo{pages}{857--872}.
\bibitem[{Diggle(1985)}]{diggle1985kernel}
\bibinfo{author}{P.~Diggle},
\newblock \bibinfo{title}{A kernel method for smoothing point process data},
\newblock \bibinfo{journal}{Journal of the Royal Statistical Society: Series C
  (Applied Statistics)} \bibinfo{volume}{34} (\bibinfo{year}{1985})
  \bibinfo{pages}{138--147}.
\bibitem[{url(2021)}]{url}
\bibinfo{title}{{Mississippi and Missouri River Alluvial Aquifer}},
  \bibinfo{howpublished}{Missouri Department of Natural Resources},
  \bibinfo{year}{accessed January 2021}. \URLprefix
  \url{https://dnr.mo.gov/geology/wrc/groundwater/education/provinces/riveralluviumprovince.htm}.
\bibitem[{Prior et~al.(2003)Prior, Boekhoff, Howes, Libra, and
  VanDorpe}]{prior2003iowa}
\bibinfo{author}{J.~C. Prior}, \bibinfo{author}{J.~L. Boekhoff},
  \bibinfo{author}{M.~R. Howes}, \bibinfo{author}{R.~D. Libra},
  \bibinfo{author}{P.~E. VanDorpe},
\newblock \bibinfo{title}{{Iowa's Groundwater Basics: A geological guide to the
  occurence, use, and vulnerability of Iowa's aquifers}}
  (\bibinfo{year}{2003}).

\end{thebibliography}
 
\end{document}